   \documentstyle[nato,epsf]{crckapb}

\begin{opening}
\title{Scaling Laws for Fluxon Formation in \protect\\
Annular Josephson Tunnel Junctions}

\author{R. MONACO}
\institute{ Istituto di Cibernetica del C.N.R,
           I-80078, Pozzuoli (Na), and INFM-Dipartimento di Fisica, Universita' di Salerno,\\
I-84081 Baronissi (Sa), Italy}
\author{R. J. RIVERS}
\institute{ Blackett Laboratory, Imperial College, \\
London SW7 2BZ}

\end{opening}

\runningtitle{FLUXON FORMATION IN JTJs}

\begin{document}


\section{Scaling Laws for Fluxon Production}
\subsection{BACKGROUND}
Although equilibrium, or adiabatic, correlation lengths
$\xi_{ad}(T)$ diverge at the critical temperature $T_c$ of
continuous phase transitions, correlation lengths always remain
bounded, in practice. This is because causality prevents a system
becoming ordered on very large scales within the finite time in
which transitions are implemented. In consequence, the order
parameter fields become frustrated, and defects arise to mediate
between the different equivalent ground states of the system. By
observing these defects (in our case, $\it fluxons$) we obtain a
direct experimental guide to the way in which the transition has
been implemented.

Causal bounds on defect densities produced in the early universe
at GUT transitions were proposed by Kibble\cite{kibble1,kibble2}
but, because of uncertainty in the models, these predictions are
not robust. However, Zurek suggested\cite {zurek1,zurek2} that
identical causal arguments were applicable to condensed matter
systems, for which direct experiments on defects could be
performed.

The reasoning is that, as the temperature $T(t)$ falls through
$T_c$ (at time $t=0$, say), there is a maximum speed $c(t) =
c(T(t))$ at which the system can become ordered. Causality
suggests that a domain structure, with defects linking domains,
cannot be established before time ${\bar t}$, determined by
$|{\dot\xi}_{ad}({\bar t})| = c ({\bar t})$, the first time at
which the rate of collapse of the adiabatic correlation length is
comparable to the maximum permitted speed. For times $t<{\bar t}$
the system is  frozen, and ${\bar t}$ measures the first time that
we re-enter an adiabatic
regime\cite{kibble1,kibble2,zurek1,zurek2}. In appropriate units
${\bar t} = t_0 (\tau_Q /\tau_0)^{\nu}$, where the  exponent $\nu$
depends on the system, and $\tau_Q$ is the quench time.

The second ingredient of the Kibble-Zurek analysis was to identify
the defect separation at their time of formation as
\begin{equation}
{\bar\xi} = \xi_{ad}({\bar t})= \xi_0 (\tau_Q/\tau_0)^{\sigma},
\label{KZ}
\end{equation}
The value of $\sigma$ depends on the system, as do $\xi_0$ and
$\tau_0$. It is this scaling behaviour that we have tested in our
experiment, using annular Josephson tunnel junctions (JTJs), for
which the defects are {\it fluxons}.

In fact, the situation is more complex. Defects can only appear
with the correct energy profiles when the order parameter has
achieved its final state magnitude. This is effected by the
exponential growth of unstable modes and is largely completed by
the spinodal time $t_{sp}$, for a transition implemented by phase
separation. However, for simple systems it can be shown
\cite{RKK,RR} that, although $t_{sp}>{\bar t}$, $t_{sp}={\bar t}$,
up to $O(1)$ multiplicative logarithmic corrections in $\tau_Q$
and the microscopic parameters of the system.

Further, for simple cases, when defects are sheets, tubes or balls
of the unstable ground state, the separation of defects is not
governed directly by $\xi_{ad}$, derived from the large-distance
behaviour of the field correlator. Rather, it is governed by the
separation $\xi_{zero}$ of the field zeroes that label these false
groundstates, determined by the short-distance behaviour of the
correlation functions. We shall see that this very different
understanding of how defects form is crucial for our analysis,
when we consider systems of linear size $L<{\bar\xi}$.
Nonetheless, the differences between $\xi_{ad}({\bar t})$ and
$\xi_{zero}(t_{sp})$ still remain logarithmic in $\tau_Q$, and the
simple scaling behaviour (\ref{KZ})  is reliable at the level of
accuracy of the experiments \cite
{helsinki,grenoble,lancaster,lancaster2,technion,pamplona,florence}
that, prior to ours, have been performed to check it.

\subsection{The Scaling Predictions For Fluxons}

The order parameter of a Josephson tunnel junction at temperature
$T<T_{c}$ is the phase difference $\phi $ of the macroscopic
superconducting quantum mechanical wave functions across the
barrier.  For an annular JTJ (AJTJ) with a bias current $\Gamma $,
$\phi $ obeys \cite{lom} the  perturbed Sine-Gordon equation
\begin{equation}
\frac{\partial ^{2}\phi }{\partial x^{2}} -\frac{1}{c^{2}(T)}\frac{%
\partial ^{2}\phi }{\partial t^{2}}-\frac{1}{\lambda _{J}^{2}(T)}\sin \phi
=\Gamma +\frac{\alpha }{c^{2}(T)}\frac{\partial \phi }{\partial
t}-\beta \frac{\partial ^{3}\phi }{\partial x^{2}\partial t}
\label{SG}
\end{equation}
provided the width $\Delta r$ of the annulus, of radius $r$, satisfies $%
\Delta r\ll r$ and $\Delta r\ll \lambda _{J}(T)$, the Josephson
coherence length; $x$ measures the distance along the annulus,
$c(T)$ is the Swihart velocity., and $\alpha $ and $\beta $ are
the coefficients of the losses due to the tunneling current and to
the surface impedance, respectively.

The boundary conditions for Eq.(\ref{SG}) are periodic, $\phi
(x+C)=\phi (x)+2\pi n$, where $C=2\pi r$ is the circumference of
the junction and the winding number $n$ is an integer
corresponding to the algebraic sum of fluxons trapped in the
junction barrier at the normal-superconducting (N-S) transition.

The classical fluxons are the 'kinks' of the Sine-Gordon theory.
Eq. (\ref{SG}) is only valid once the transition is complete, and
we shall not use it to study the appearance of fluxons. However,
it is sufficient to enable us to identify $\lambda _{J}(T)$,
diverging at $T_{c}$, as the equilibrium correlation length $\xi
_{ad}(T)$ to be constrained by causality. Further, the Swihart
velocity $c(T)$ measures the maximum speed at which the order
parameter can change\cite{Swihart,Barone}.

 For the experiment that we shall describe below,
 we find
that $C<{\bar\xi}$. In fact, because we are really counting field
zeroes (mod $2\pi$), the conclusions are equally valid for small
systems. Any field crossing zero (mod $2\pi$) has the potential to
mature into a fluxon. Thus, before the causal time we have a
picture in which there is a fractal thermal fuzz of potential
fluxons, whose density depends on the scale at which we look. By
$O({\bar t})$ some of these have developed into the fluxons that
we see subsequently (see \cite{RKK,RR}). However, because from
this viewpoint we have many proto-defects jockeying to become the
real thing, the relevant scale to compare to the system size is
not ${\bar\xi}$ but $\xi_0$, and $C\gg\xi_0$.

 A detailed discussion of JTJs has been given elsewhere by us \cite {KMR,MRK},
 including a summary of this experiment \cite {MMR}, and we refer the
reader to these articles for more details. The JTJs in our
experiment are {\it symmetric}, i.e. the electrodes are made of
identical superconducting material. For such JTJs  $\nu =1/2$ and
$\sigma = 1/4$. Therefore, at the time of their formation the
separation of fluxons is expected to be
\begin{equation}
{\bar{\xi}}\sim \xi _{0}(\tau _{Q}/\tau _{0})^{1/4}. \label{JTJ}
\end{equation}
In terms of the parameters of the JTJs, $ \xi _{0}=\sqrt{\hbar
/2e\mu _{0}d_{s}aJ_{c}(0)} $, where $J_{c}(T)$ is the critical
Josephson current at temperature $T$ for a JTJ with
superconductors with effective thickness $d_s$. The parameter $a$
is given in terms of the gap energy and critical temperature and
has a value between $3$ and $5$.  As for $\tau _{0}$, it is given
as $\tau _{0}=\xi _{0}/c_{0}$, where $c_{0}$ defines the behavior
$c(t)=c_{0}(t/\tau _{Q})^{1/2}$of the Swihart velocity for the
system near $T=T_{c}$.

\section{The Experiment}
\subsection{Measuring Fluxons}

When fluxons are produced they are static. To see them it is
necessary to feed a bias current to the AJTJ, whereupon they move
as magnetic dipoles. As a result, they leave a clear signature on
the junction current-voltage characteristic (IVC). Fluxons having
different topological charge $n=\pm 1$ travel in opposite
directions. Quantitatively, if a fluxon travels around an AJTJ
having circumference $C$ with a constant speed $v<c(T)$, then an
average voltage $V=\Phi _{0}v/C$ develops across the junction. By
changing the bias current through the barrier the voltage drop
changes and a new branch  appears on the junction IVC. When $N$
fluxons travel around an AJTJ, the junction voltage is $V=N\Phi
_{0}C/v$. In this expression $N$ is the total number of travelling
fluxons and can be larger than the winding number $n$ if
$F\overline{F}$ pairs are travelling around the annulus.
Therefore, we can count the number of travelling fluxons by simply
measuring the voltage across the AJTJ.

To show this, Figs.1 and 2 represent the IVC of the same AJTJ with
no fluxon trapped and with one fluxon trapped, respectively.
\begin{figure}
{\centerline{\epsfysize=5cm   \epsffile{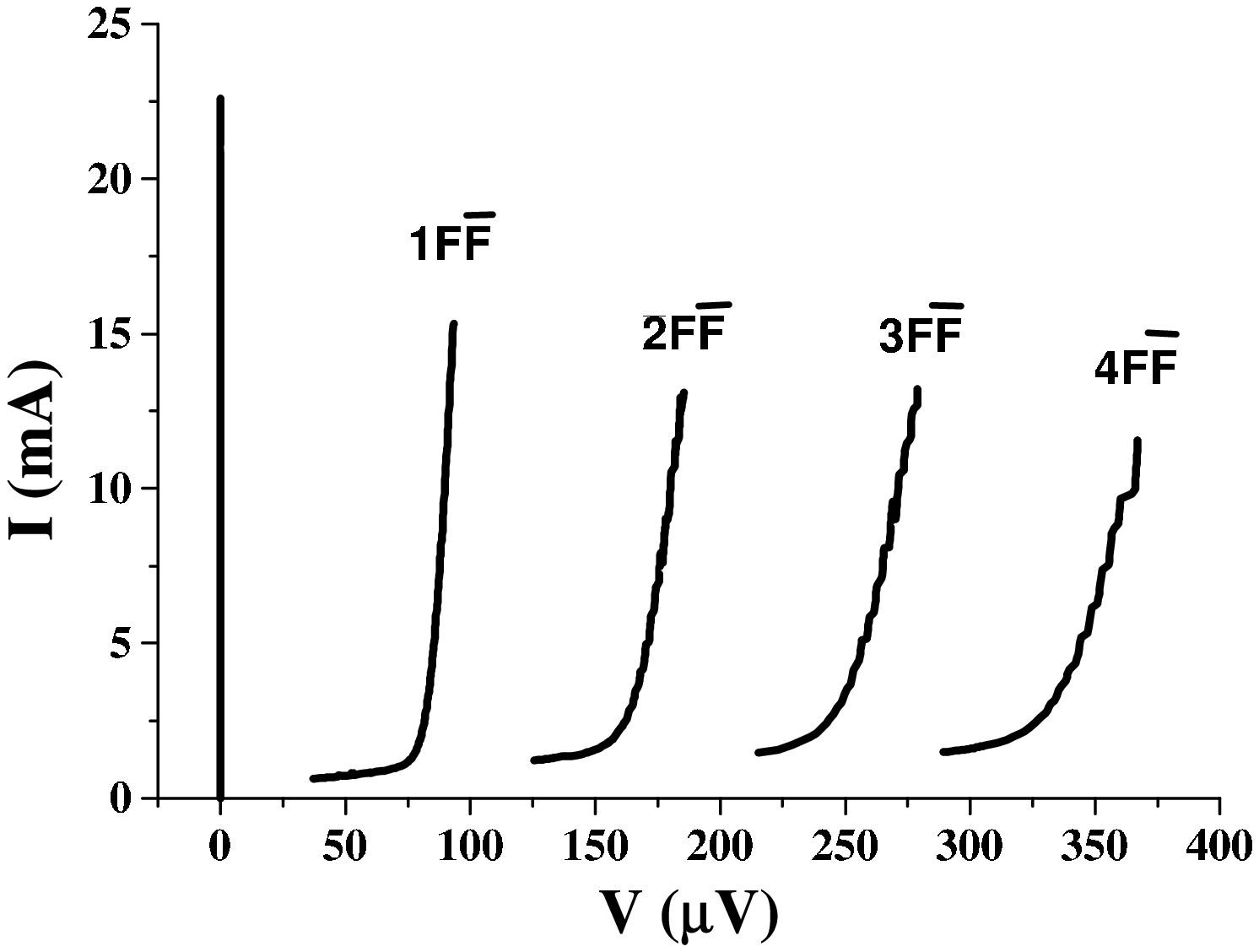}}}
 \caption{The measured current-voltage characteristics of an AJTJ
without trapped fluxons. For each current branch the number of
fluxon-antifluxon pairs $F{\bar F}$ is indicated.}
\end{figure}
\begin{figure}
{\centerline{\epsfysize=5cm   \epsffile{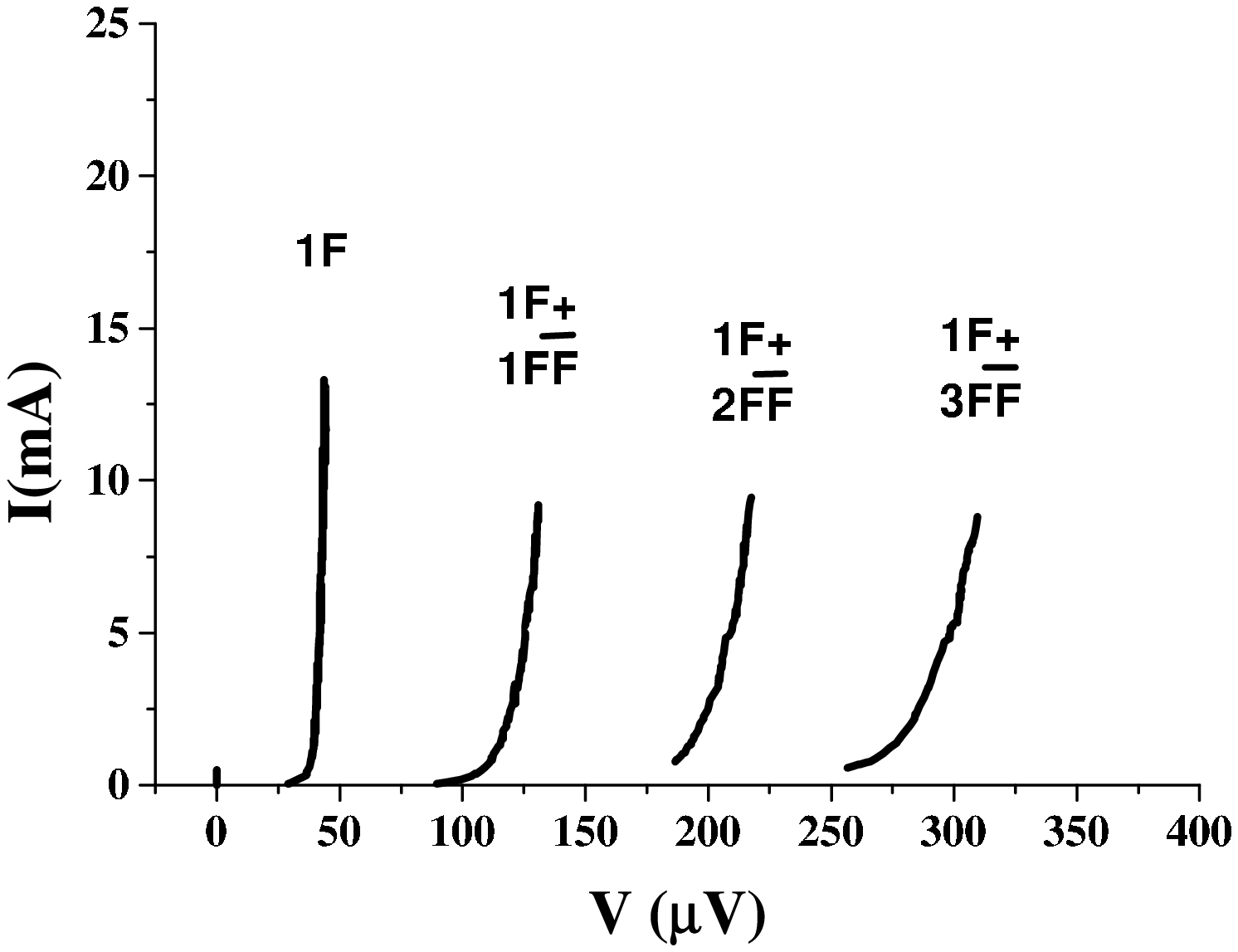}}}
\caption{The current-voltage characteristics of the same annular
JTJ as in Fig.1 with one trapped fluxon and fluxon-antifluxon
pairs.}
\end{figure}
We note that with no trapped fluxons the zero voltage current is
very large. In the other case the supercurrent is rather small
(ideally zero) and large current branches can be observed at
finite voltages corresponding to the fluxon and, possibly,
$F\overline{F}$ pairs travelling around the junction.

\subsection{The experimental setup}
High quality $Nb/Al-Al_{ox}/Nb$ JTJs were fabricated on $0.5\,mm$
thick silicon substrates (chips) using the trilayer technique
described in Ref.\cite{VPK}. Each JTJ had a mean circumference
$C=500\,\mu m$ and a width $\Delta r=4\,\,\mu m$. The
thicknesses of the base, top and wiring layer were $200$, $80$ and $400\,nm$%
, respectively. The high quality of the samples was inferred from
the I-V characteristic at $T=4.2\,K$ by checking that the subgap
current $I_{sg}$ at $2\,mV $ was small compared to the current
rise $\Delta I_{g}$ in the quasiparticle current at the gap
voltage $V_{g}$.

The symmetry of the junctions is assured by the absence of a
logarithmic singularity in the IVCs at low voltages and  the
linear temperature dependence of the critical current as the
temperature $T$ approached the critical temperature $T_{C}$. Many
samples have been measured. For clarity only two will be discussed
here. Their geometrical and electrical (at $4.2\,K$) parameters of
are listed in Table I. In particular, they
differ in their critical current densities. The  Swihart velocity $%
c_{0} $ has the value $c_{0}=1.4\times 10^{7}m/sec$.

\begin{table}[htb]
\begin{center}
\caption{The samples}
\begin{tabular}{lll}
\hline Sample & A & B\\
\hline
Mean circumference $C(\mu m)$ & 500 & 500 \\
Width $\Delta r(\mu m)$ & 4 & 4 \\
Zero field critical current $I_{o}(mA)$ & 33 & 2.5 \\
Maximum critical current $I_{\max }(mA)$ & 39 & 2.7 \\
Gap quasiparticle current step $\Delta I_{g}(mA)$ &
88 & 5.2 \\
$I_{\max }/\Delta I_{g}$ & 0.45 & 0.52 \\
Critical current density $J_{c}(A/cm^{2})$ & 3050 & 180 \\
Josephson length $\lambda _{J}(\mu m)$ & 6.9 & 28 \\
Normalized mean circumference $C/\lambda _{J}$ & 72 & 18 \\
Quality factor $V_{m}(mV)$ & 49 & 63 \\
Normal resistance $R_{N}(m\Omega )$ & 36 & 610 \\
ZFS1 asymptotic voltage $(\mu V)$ & 51 & 53 \\
\hline
\end{tabular}
\end{center}
\end{table}

In order to vary the quenching time in the broadest possible
range, we have realized the experimental setup schematically shown
in Fig.3.
\begin{figure}
{\centerline{\epsfysize=7cm   \epsffile{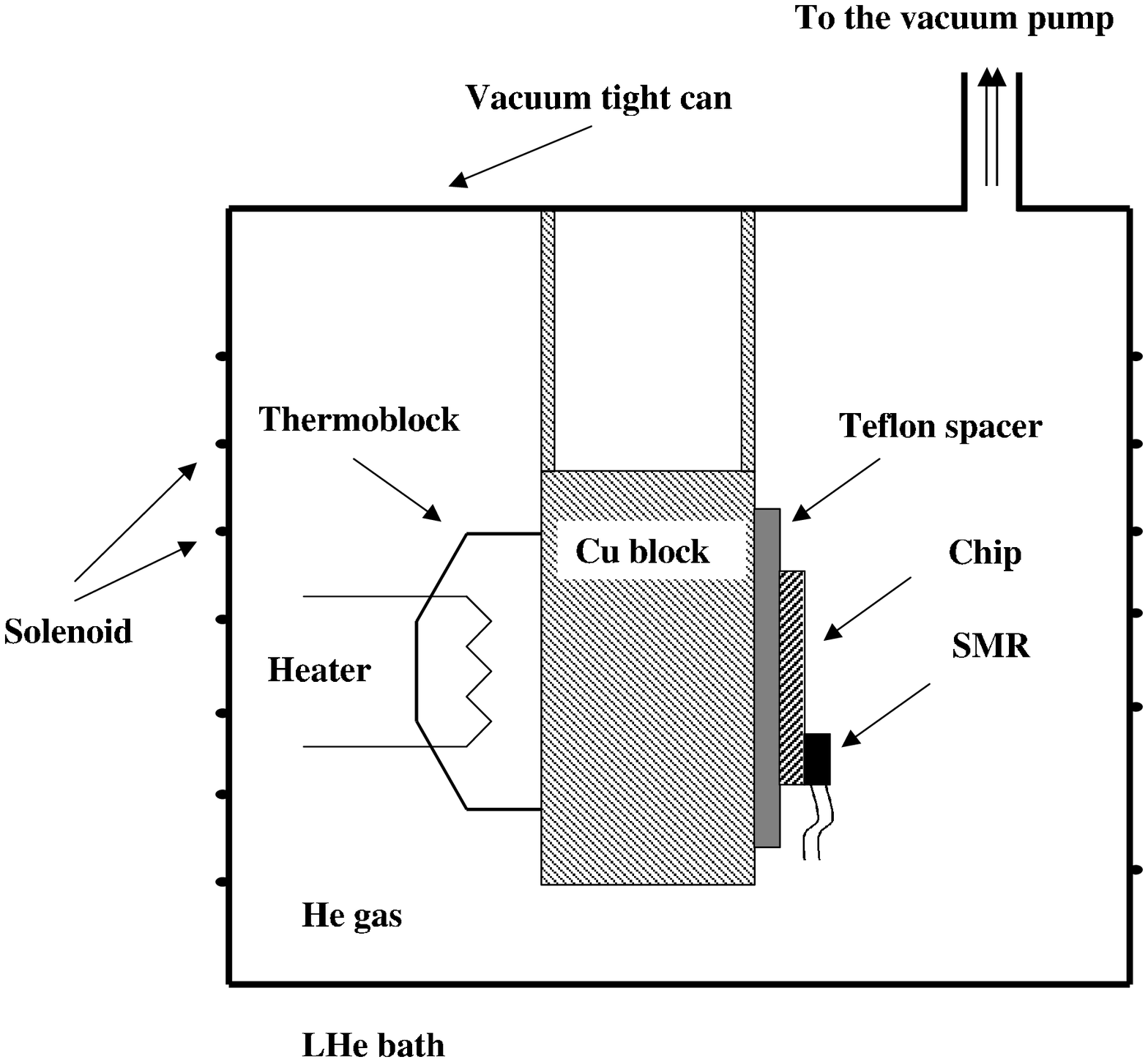}}}
\caption{Sketch (not to scale) of the cryogenic insert developed
to perform the junction thermal cycles over slow and fast
timescales}
\end{figure}
A massive $Cu$ block held to the sample holder  was used to
increase the system thermal capacity. The chip was mounted on one
side of this block. On the other side, a thermoblock was mounted
consisting of a $50\,\Omega $ carbon resistor and two thermometers
in order to measure and to, if necessary, stabilize the $Cu$ block
temperature. Finally a small sized $100\,\Omega $ surface mounted
resistor (SMR), was kept in good thermal contact with the chip.

This system allowed us to perform the sample quenching over two
quite different time scales. By means of the resistor in the
thermoblock, a long time scale was achieved by heating the chip
through the $Cu$ block; on the contrary, a short current pulse
through the SMR on the chip, attained much short thermal cycles.
The total system was kept in a vacuum-tight can immersed in the
$LHe$ bath at $He$ gas pipeline pressure. By changing the exchange
gas pressure and using the two techniques, the quenching time can
be changed over a quite large range from tenths to tens of
seconds.

 The temperature
dependence of the junction gap voltage was exploited to monitor
the temperature of the junction itself during the thermal cycle.
The analytical expression \cite{thouless} for the superconductor
gap energy $\Delta (T)$,

\begin{equation}
\frac{\Delta (T)}{\Delta (0)}=\tanh \frac{\Delta (T)}{\Delta (0)}\frac{T_{c}%
}{T},  \label{gap}
\end{equation}

\noindent also applies to the junction gap voltage that is
proportional to it. In this way we derive temperature profiles for
slow and fast cycles, assuming for $\Delta (0)$ and $T_{c}$
the values $2.85\,meV$ and $8.95\,K$, respectively, as found by Monaco {\it %
et al.} \cite{cristiano} on similar JTJs.

We are only interested to the cooling process, and we use
(\ref{gap}) to fit our data by a simple thermal relaxation
equation:

\begin{equation}
T(t)=T_{fin}+(T_{in}-T_{fin})\exp \left( -\frac{t-t_{0}}{\tau
}\right) \label{relaxation}
\end{equation}

\noindent with only two fitting parameters $t_{0}$ and $\tau $, and $%
T_{in} $ and $T_{fin}$ fixed at $8.95$ and $4.15\,K$, respectively. In Eq.%
\ref{relaxation} $t_{0}$ is the time at which $T=T_{in}=T_{c}$ and
$\tau $ is the relaxation time which sets the cooling time scale.
 The fitting curve for a fast
quench  is shown by a thick dashed line in Fig.4, corresponding to
a thermal relaxation time $\tau $ of $0.073\,s$. The dashed
horizontal line indicates the temperature threshold below which
the temperature time dependence can be reliably accounted for by
our measured data.
\begin{figure}
{\centerline{\epsfysize=5cm   \epsffile{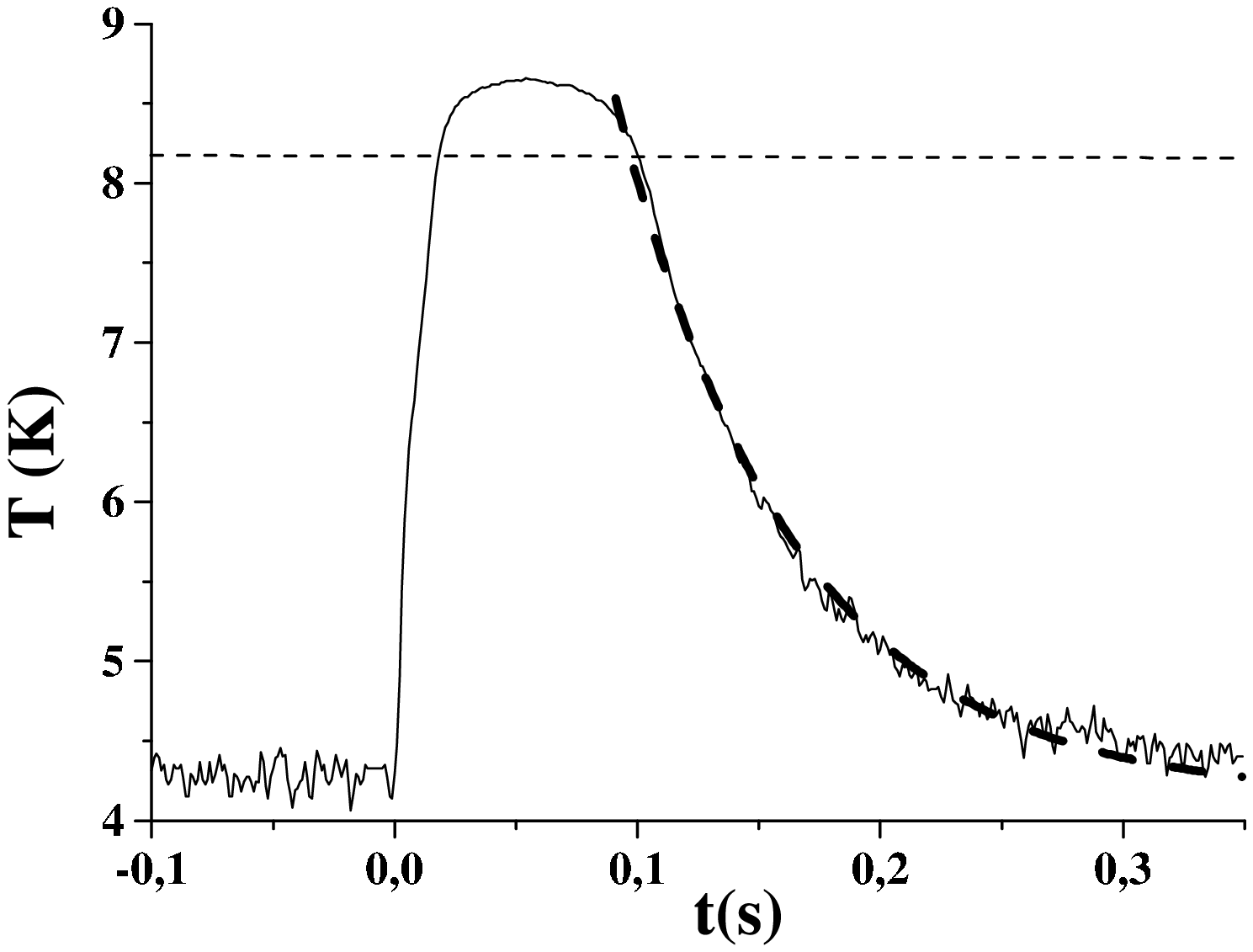}}}
\caption{Time dependence of the junction temperature (only
reliable below the horizontal dashed line) during a slow thermal
cycle, as determined from (\ref{gap}). The thick dashed line is
the best fitting curve to (\ref{relaxation}). }
\end{figure}
The quenching time $\tau _{Q}$ can be obtained from its definition
$\tau _{Q}=-T_C/{\dot T}$ at $T=T_C$, giving $\tau _{Q}=\tau
T_{C}/(T_{in}-T_{fin})$.

\subsection{The measurements}

Quenching experiments were carried out in a double $\mu $-metal
shielded cryostat and the transitions from the normal to the
superconducting states were performed with no current flowing in
the heaters and the thermometers. Both the junction voltage and
current leads were shorted during the all thermal cycle.
Furthermore, the heat supplied to the sample was such that the
maximum temperature reached by the junction was made slightly
larger than its critical temperature, say at about $10\,K$, in
order to make sure that also the bulk electrode critical
temperature ($T_{C}\simeq 9.2K$) was exceeded. In this case,
according to Eq.\ref{relaxation}, the value of the quenching time
is $\tau _{Q}\simeq 1.7\,\tau $. The value of the quenching times
has been determined to an overall accuracy of better than $5\%$.
For each value of the quenching time, in order to estimate the
trapping probability, we have carried out a set of 300 thermal
cycles and at the end of each cycle the junction IVC was inspected
in order to ascertain the possible spontaneous trapping of one or
more fluxons.

As we shall see later, the AJTJs are such that the ZK causal length ${\bar\xi%
}> C$ by roughly an order of magnitude. Thus the probability of
finding a single fluxon is small. In the following we will focus
our attention only on the probability $P_{1}$ to trap just one
fluxon, although a few times we found clear evidence of two and,
more seldom, three homopolar fluxons spontaneously trapped during
the N-S transition. However, these events were too rare to be
statistically significant. Experimentally, we define $P_{1}$ as
the fraction of times in which at the end of the thermal cycle the
junction IVC looks like Fig.1b, i.e. with a tiny critical current
and a large first ZFS. Our definition of $P_{1}$ is reasonable as
far as the chance to trap two fluxons is negligible.

\section{The results}

When ${\bar\xi}>C$ we estimate the probability of finding a fluxon
in a single quench to be
\begin{equation}
P_1 \simeq {C\over{\bar\xi}} = {C\over\xi_0} \bigg
({\tau_Q\over\tau_0}\bigg)^{-\sigma}, \label{P1}
\end{equation}
where, from (\ref{JTJ}), $\sigma = 0.25$.

Fig.5 shows on a log-log plot the measured probability $P_{1}$ of
a single fluxon trapping obtained by quenching the sample $A$ 300
times for each value of the quenching time $\tau _{Q}$ changed by
varying the exchange gas pressure and by using both the fast and
slow quenching techniques.
\begin{figure}
{\centerline{\epsfysize=5cm   \epsffile{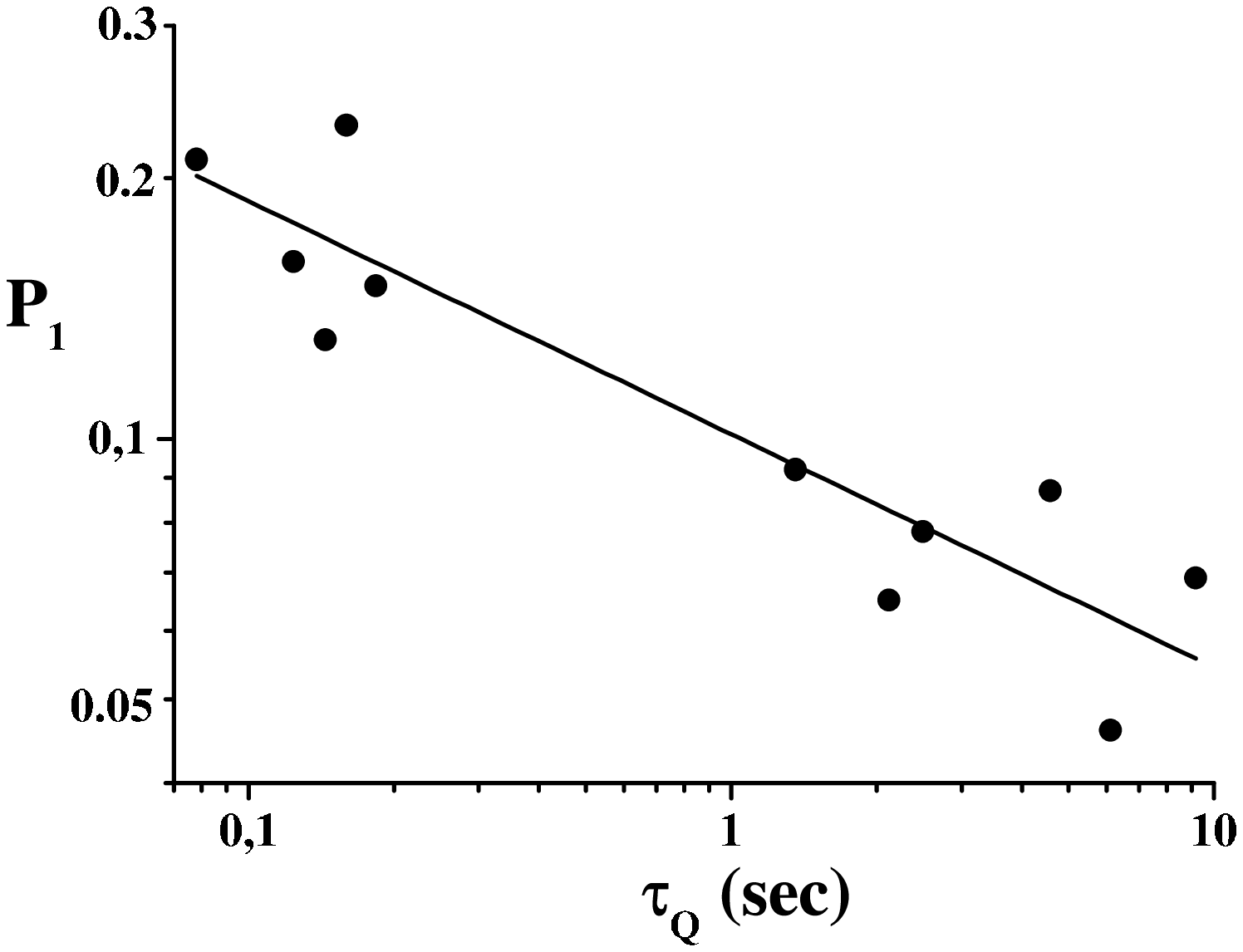}}}
\caption{Log-log plot of the probability $P_1$ to trap one fluxon
versus $\tau_Q$. The solid line is the best fitting curve,
assuming a power law dependence as in (\ref{KZ}). }
\end{figure}
Although the points are quite scattered, we attempted to fit the
data with an allometric function $P_{1}=a\cdot \tau _{Q}^{-b}$
with $a$ and $b$ being free fitting parameters. We found that the
best fitting curve, shown by the solid line in Fig.6, has a slope
$b=0.27\pm 0.05$. Such a value of $b$, although affected by a
$20\%$ uncertainty, is in good agreement with the expected
$b=0.25$ dependence.

For the coefficient $a$ we found the best fitting value of $0.1\pm 10\%$ ($%
\tau _{Q}$ in seconds). This is to be compared with the predicted value of $%
C\tau _{0}^{1/4}/\xi _{0}\,$. Sample $A$ had a circumference
$C=500\,\mu m$. Its effective superconductor thickness was
$d_{s}\approx 250$ $nm$. At the
final temperature $T_{fin}=4.2\,K$, the critical current density was $%
J_{c}(T_{fin})=3050\,A/cm^{2}$ and the Josephson length was
$\lambda _{J}(T_{fin})=6.9\,\mu m$. From this, and $c_{0}$ given
earlier, we infer that $\xi _{0}\approx 3.8\,\mu m$ and $\tau
_{0}\approx 0.17\,ps$. This then gives $C\tau _{0}^{1/4}/\xi
_{0}\approx 0.08\,s^{1/4}$, in good agreement with the
experimental value of $a$, given the fact that we only expect
agreement in overall normalization to somewhat better than an
order of magnitude level.

Similar measurements have been carried out for sample $B$.
Although compatible  with (\ref{JTJ}), the results were affected
by a data scattering even larger than that found for sample $A$.
This is due to a much smaller normalized length which, according
to (\ref{P1}), translates in a expected probability $P_{1}$, for a
given $\tau _{Q}$, about four times smaller, far too small to get
statistically significant data in reasonable times.  However the
roughly measured probability $P_{1}$ of 1 fluxon every 50-100
attempts is in fairly good aggreement with the expected value.

\section{Comments, Future Experiments and Conclusions}

We consider this experiment to give a striking confirmation of the
Zurek-Kibble predictions.

Of the several experiments [7-13] that had been performed prior to ours to test (%
\ref{KZ}) only  two \cite{pamplona,florence} had variable $\tau
_{Q}$ and could confirm scaling behaviour.
 Of these, only one
experiment \cite{florence} shows KZ scaling. Given this relatively
poor success rate in confirming (\ref{JTJ}) we are considering a
further experiment to measure the ZK scaling exponent, this time
with manifestly non-symmetric AJTJs, which require different
fabrication techniques. In this case the value of $\sigma $
inferred from the same causal arguments is $\sigma =1/7$, very
different from the $\sigma =1/4$ that we tested above.

Our experiments have demonstrated that quenching times of the
order of 1 second give a rather large probability to trap one
fluxons on AJTJs having a very large normalized length. However,
very long junctions mean very large critical current densities
that, in turn, require Josephson barriers so thin that their
quality and uniformity is often spoiled. For these reasons, it
would be highly desirable to compensate the reduced junction
length with an increased quenching rate, as it is suggested by the
findings for sample $B$.
 On the contrary, in order to lower the
quenching time,  it is necessary to resort to new techniques since
the maximum power that can be dissipated by the surface mount
resistors sets an obvious lower threshold on $\tau _{Q}$.
\smallskip One possible way to reach this goal is to perform the
junction thermal cycle by means of light pulses. Light dissipates
inside the superconducting electrodes, but not in the substrate
providing a local junction heating that will relax much faster to
the background temperature. We estimate that, by using a properly
focussed pulsed light beam, the quenching time scale can be
reduced to the $\mu s$ range.

\section*{Acknowledgements}
The authors thank J. Mygind for invaluable help in conducting the
experiment and L. Filippenko for the sample fabrication.

\end{document}